\documentclass[oldversion,paper]{aa} 

\usepackage{amsmath,texlogos} 
\usepackage{natbib}
\usepackage{subfigure}
\usepackage{graphicx}
\usepackage{txfonts}

\begin{document}

\titlerunning{Transiting exoplanet masses from precise optical photometry}
\authorrunning{D. Mislis et. al.}

  \title{Estimating transiting exoplanet masses \\ from precise optical photometry}

  \author{D. Mislis \inst{1}, R. Heller \inst{2}, J.H.M.M. Schmitt \inst{3}, S. Hodgkin \inst{1}}  

\institute{$^{1}$Institute of Astronomy, Madingley Road, Cambridge CB3 0HA, UK\\
$^{2}$Leibniz-Institut f\"{u}r Astrophysik Potsdam (AIP), An der Sternwarte 16, 14482 Potsdam, Germany\\
$^{3}$Hamburger Sternwarte, Gojenbergsweg 112, D-21029 Hamburg \\
             \email{misldim@ast.cam.ac.uk}
            }

\date{Accepted : 08/12/2011}

\abstract{

  We present a theoretical analysis of the optical light curves (LCs)
  for short-period high-mass transiting extrasolar planet
  systems. Our method considers the primary transit, the secondary
  eclipse, and the overall phase shape of the LC between the
  occultations. Phase variations arise from (i) reflected and
  thermally emitted light by the planet, (ii) the ellipsoidal shape
  of the star due to the gravitational pull of the planet, and (iii) the
  Doppler shift of the stellar light as the star orbits the center of
  mass of the system. Our full model of the out-of-eclipse variations
  contains information about the planetary mass, orbital
  eccentricity, the orientation of periastron and the planet's
  albedo. For a range of hypothetical systems we demonstrate that 
  the ellipsoidal variations (ii.) can be large enough to be distinguished 
from the remaining components and that this effect can be used to constrain 
the planet's mass. To detect the ellipsoidal
  variations, the LC requires a minimum precision of $10^{-4}$, which
  coincides with the precision of the Kepler mission. As a test of our
  approach, we consider the Kepler LC of the transiting object
  HAT-P-7. We are able to estimate the mass of the companion, and
  confirm its planetary nature solely from the LC data. Future space
  missions, such as PLATO and the James Webb Space Telescope with
  even higher photometric precision, will be able to reduce the errors
  in all parameters. Detailed modeling of any out-of-eclipse
  variations seen in new systems will be a useful diagnostic tool
  prior to the requisite ground based radial velocity follow-up.

}

\keywords{Stars: Planetary systems -- Methods: data analysis, analytical -- Techniques: photometric -- Celestial mechanics}

\maketitle

\section{Introduction}

Two observational methods have dominated the study of extrasolar
planets so far: radial velocity (RV) measurements and transit light
curve (LC) analyses. Both have advantages and disadvantages. While 
RV determinations provide estimates of the planetary mass
$(M_\mathrm{p})$, the eccentricity $(e)$ and the semi-major axis
$(a)$, they do not constrain the inclination $(i)$ of the orbital
plane with respect to the observer, thus only lower limits to
$M_\mathrm{p}$ can be determined. The transit method, on the other
hand, provides information on $i$, the ratio of the planetary to the
stellar radius $(R_\mathrm{p}/R_\mathrm{s})$, and the duration of the
transit $(D)$. So far, only a combination of both strategies yields a
full set of orbital and physical parameters for extrasolar
planets.\par

Currently, two space-missions are targeted at the detection of
transiting extrasolar planets: CoRoT launched in 2006
\citep{1997ASPC..119..259D} and Kepler launched in 2009
\citep{1997ASPC..119..153B}. Their instruments are monitoring hundreds
of thousands of stars, discovering thousands of planet candidates,
whose RV follow-up will take many years (and may never be completed
for the faintest candidates). Without RV follow-up, the most
fundamental parameter of an extrasolar planet, its mass, remains
undetermined from CoRoT and Kepler observations alone. The mass is the
crucial parameter classifying an object as a planet, brown dwarf or a
star. \par

High-accuracy photometry has already been used for a number of systems
to show that the planetary thermal emission, as well as the reflection
of the stellar light from the planet, are detectable. In particular, Welsh et
al. (2010) report the discovery of ellipsoidal variations in the
Kepler LC of HAT-P-7. This is an effect more commonly known from close 
stellar systems, where phase-dependent light variation
arises from the gravitationally distorted stars. In HAT-P-7, 
the planet is close enough and massive enough to induce the same effect.

Other authors \citep{2003ApJ...588L.117L, 2011MNRAS.415.3921F} have already presented treatments of reflected light, thermal emission, ellipsoidal variations and Doppler beaming. Our approach to the problem includes two main differences: (i) we investigate the sensitivity of our technique on the adopted phase function (Section 3.2), and (ii) we attempt to solve separately for the reflected light and thermal emission from the planet. For hot Jupiters around hot stars, the temperature of the planet increases dramatically and reflected light alone fails to explain the LC.

In this paper, we investigate whether the presence of ellipsoidal
variations in a planetary system can be used to place meaningful
constraints on the mass of the planetary companion. In
Sect. \ref{sec:theory} we describe the basic phenomena included in
our simplified model of a planetary system's LC. In
Sect. \ref{results} we compare the magnitudes of the relative
contributions and consider the likely range of planetary systems for which
ellipsoidal variations ought to be detectable with both CoRoT and
Kepler. We reanalyze the HAT-P-7 LC and demonstrate that the
ellipsoidal variations can be used to estimate the mass of the planet
to within 10\% of the radial-velocity measured mass. We summarize our
conclusions in Sect. \ref{conclusions}.

\section{Parametrization of photometric transits}
\label{sec:theory}

\begin{figure}[h]
  \centering
  \includegraphics[width=8.7cm]{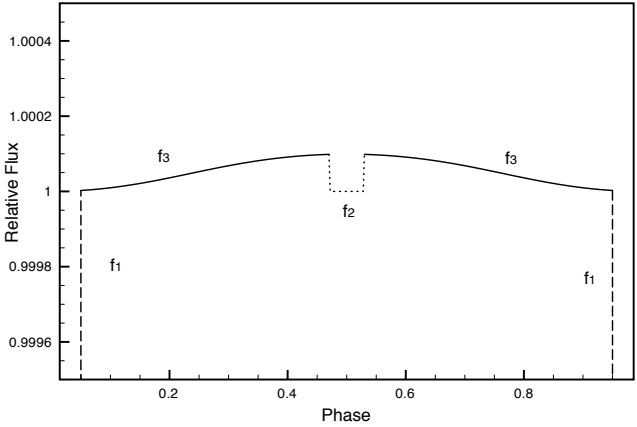}
  \caption{The standard transit model includes the primary transit
    $f_1$ (dashed line centred at orbital phase $z = 0$), the
    secondary eclipse $f_2$ (dotted line around $z = 0.5$) and the
    reflected light from the planet $f_3$ (solid line). The model is
    based on a HAT-7-b twin planet (Table \ref{tab2} ).}
  \label{fig1}
\end{figure}

Standard models of LCs that have been used before the advent of space
missions are based on a flat curve out of transit and a limb darkening
during the transit. \citet{2003ApJ...585.1038S} proved analytically
that for each package of planet-star characteristics there is a unique
LC. Analyses of high-accuracy data from space required a revision of
this simple approach. For a high accuracy LC, the model should
incorporate the reflected light from the planet, thermal emission,
ellipsoidal variations and Doppler boosting, which deforms the overall
shape of the LC, and the secondary eclipse (Fig.~\ref{fig1}).

\noindent

\begin{figure}[h]
  \centering
  \includegraphics[width=6.cm]{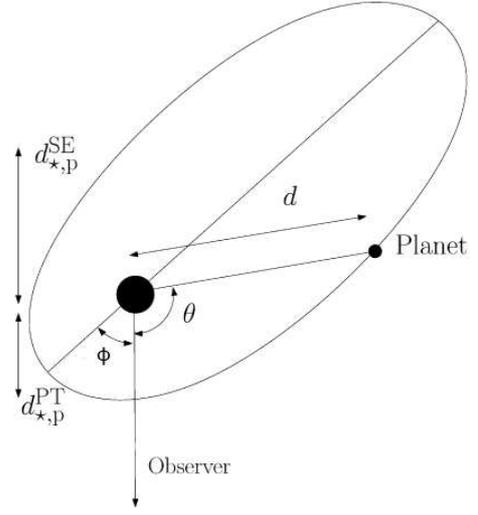}
  \caption{This sketch of the transiting planetary system, as seen from above, explains the variables used in our calculations. We assume 
clockwise rotation.}
\label{fig2}
\end{figure}

In Fig.~\ref{fig2} we show the geometry of an arbitrary transiting
system assuming an elliptical orbit. Let $i$ be the angle between the
observer's line of sight and the orbital plane normal, while the angle
between the observer's line of sight projected onto the orbit plane
and the periastron is labeled $\phi$. The star is in the center of
the reference frame and $d$ is the distance between the star and the
planet; the distance between the star and the planet during the
primary transit is denoted by $d_{\star,\mathrm{p}}^{\mathrm{PT}}$,
during the secondary eclipse both bodies are separated by the distance
$d_{\star,\mathrm{p}}^{\mathrm{SE}}$.

To decode the geometry of the system from the LC we split it into
three sub LCs. LC $f_1$ describes the course of the primary
transit, when the planet blocks the star's light, $f_2$ describes the
form of the secondary eclipse, when the star blocks and planetary
light, and $f_3$ the rest of the LC, when both star and
planet contribute to the total amount of light. \par

\subsection{Transit and Eclipse}

The duration of the transit $f_1$ is given by

\begin{equation}
D \simeq  \displaystyle\frac{PR_\star}{\pi a_\mathrm{d}^{2} }\frac{d_{\star,\mathrm{p}}^{\mathrm{PT}}}{\sqrt{1-e^{2}}}\sqrt{ \left( 1+\frac{R_\mathrm{p}}{R_\star} \right)^{2}-b^{2}} ,
\label{eq1}
\end{equation}
\noindent

where $a_\mathrm{d}$ is the semi-major axis of the system, $P$ is the
orbital period, $R_\star$ and $R_\mathrm{p}$ are the radius of the
star and the planet, respectively, $e$ is the orbital eccentricity,
$b=d_{\star,\mathrm{p}}^{\mathrm{PT}}\cos i/R_\star$ is the impact
parameter \citep{2008ApJ...678.1407F} and $i$ is the inclination of
the orbital plane with respect to the observer's line of sight. To
model the shape of $f_1$ we use Eq. (\ref{eq1}) and the limb darkening
equation

\begin{equation}
\frac{I_\mathrm{\mu}}{I_0} = 1 - u_1(1-\mu) - u_2(1-\mu)^{2}, \label{eq2}
\end{equation}
\noindent

with $u_1$ and $u_2$ the linear and quadratic limb darkening
coefficients respectively \citep{2004A&A...428.1001C,
  2007ApJ...664.1190S}, $\mu$ as the cosine of the angle between the
surface normal and the observer, and $I_0$ and $I_{\mu}$ as the
intensities at the stellar disk center and at $\mu$,
respectively. Once the period is known from observations, one can fit
the model to the observations to deduce $R_\star$, $R_\mathrm{p}$,
$i$, and $d_{\star,\mathrm{p}}^{\mathrm{PT}}$. The transit of the
secondary eclipse, $f_2$, is fitted with the same model but without
the effects of limb darkening.

\subsection{Phase-dependent light curve}

The shape of the LC between the transit and
the eclipse ($f_3$) contains five contributions: (i) $f_\star$, a
contribution from the stellar photospheric flux, which almost
certainly varies with time, but which we assume to be constant
throughout this analysis; (ii) $f_\mathrm{ref}$, a contribution from
starlight reflected from the face of the planet, which is phase
dependent; (iii) $f_\mathrm{th}$ contributions from the intrinsic
planetary thermal emission (split by day and night), which depends
sensitively on a large number of factors, including the spectral range
of the observation $\lambda$, the irradiation temperature, and the
atmospheric composition and structure; (iv) $f_\mathrm{ell}$, a
contribution from ellipsoidal variations of the star itself due to
tidal forces from the planet; (v) flux variations $f_\mathrm{dop}$ due
to a Doppler shift (Doppler boosting) of the stellar spectral energy
distribution with respect to the bandpass of the instrument
\citep{2010A&A...521L..59M}. We normalize the LC by the stellar flux,
which can be determined as the minimum flux observed at phase $z
= 0.5$. The phase LC then becomes

\begin{equation}
f_3(z) = \frac{f_\star + f_\mathrm{ref} + f_\mathrm{th,d}+f_\mathrm{th,n}+f_\mathrm{ell}+f_\mathrm{dop}}{f_\star}. \label{eq8}
\end{equation}
\noindent

In the following, we study each of these terms in more detail.

\subsubsection{Reflected light}

The phase pattern of the reflected stellar flux depends on the phase
angle $z$, i.e., the angle between star and observer as seen from
the planet. Counting the orbital phase $\theta$ from primary minimum,
the angles $z$, $i$ and $\theta$ are related through

\begin{equation}
\cos(z)=  -\sin(i)\cos(\theta).
\label{eq33}
\end{equation}
\noindent

The reflected flux $f_\mathrm{ref}$ can then be expressed as

\begin{equation}
f_\mathrm{ref}(z) = \alpha_\mathrm{g} f_\star  \left( \frac{R_\mathrm{P}}{d} \right)^{2} \Phi (z),
\label{eq3}
\end{equation}
\noindent
where $\alpha_\mathrm{g}$ is the geometric albedo of the planet,
$f_\star$ is the stellar flux at a distance $d$ from the star, and
$\Phi (z)$ is the so-called phase function. It is not entirely
clear which phase function provides the most appropriate description
of extrasolar planets. A popular assumption is
\begin{equation}
\Phi_{L}(z) = \frac{1}{\pi} {\Big(} \sin{z}+(\pi-z) \cos{z}  {\Big)},
\label{eq5}
\end{equation}
which models the planet as a Lambert sphere \citep{1916ApJ....43..173R}, assuming that the intensity of the reflected light is 
$\Phi_{L} = 1 / \pi$ at at phase $z= \pi / 2$ (Venus case). An alternative choice is 
\begin{equation}
\Phi_{C}(z) = \frac{1}{2} {\Big(} 1+\cos{(z)} {\Big)},  
\label{eq4}
\end{equation}
which assumes that the reflected light is $\Phi_{C}=0.5$ at phase $z= \pi / 2$. In our model, we investigate both cases. \par

The sum $\phi\ + \theta$ is related to the eccentric anomaly $E$ through
\begin{equation}
\phi + \theta = 2 \tan^{-1} \left(  \frac{\sqrt{1+e}}{\sqrt{1-e}} \tan{(E/2)}  \right) ,
\label{eq7}
\end{equation}
and $E$ is related to the mean anomaly $M$ through Kepler's equation
\begin{equation}
E=M-e\cos(E). 
\end{equation}

If $e \neq 0$ then the geometrical albedo is not constant and it is given by the equation Eq. \ref{alb}  \citep{2010ApJ...724..818K}
\begin{equation}
\alpha_\mathrm{g} =   \frac{e^{d-1}-e^{1-d}}{5(e^{d-1}+e^{1-d})} + \frac{3}{10}  \label{alb}
\end{equation}

\subsubsection{Ellipsoidal variations}

When a Jupiter-like planet orbits close to its host star, say $a~\lesssim~0.1\,\mathrm{AU}$, then the star will be distorted. Hence, 
the sky-projected shape will vary, causing flux variations. Their magnitude is given by
\begin{equation}
\frac{f_{\mathrm{ell}}}{f_{\star}} = \beta    \frac{M_{\mathrm{p}}}{M_{\star}} \left( \frac{R_{\star}}{a_{\mathrm{d}}}  \right)^{3} \left( \frac{1+e\cos(\phi+\theta)}{1-e^{2}} \right)^{3} \sin^{3}(i) |\sin(\theta)| ,
\label{eq443}
\end{equation}
\noindent
where
\begin{equation}
\beta =  \frac{\log_{10} \left( G M_{\star} /  R_{\star}^{2}   \right) }{\log_{10}T_{\mathrm{eff}}} 
\label{eq555}
\end{equation}
\noindent
is the gravity darkening term (radiative energy transfer - \cite{1924MNRAS..84..665V}), and $M_{\mathrm{p}}$ and $M_{\star}$ are the masses of the planet and the star, respectively. As these equations show, the detection of the ellipsoidal distortion of the star provides information about the planetary mass with respect to the stellar mass.

\subsubsection{Doppler boosting}
Finally, the relative flux because of stellar Doppler shifts given by
\begin{equation}
\frac{f_{\mathrm{dop}}}{f_{\star}} = \left( 3-\rho \right) \frac{K}{c} ,
\label{eq444}
\end{equation}
\noindent
with $c$ as the speed of light,
\begin{equation}
K = 28.4\cdot P^{-1/3} \cdot M_{\mathrm{p}}\cdot \sin(i) \cdot M_{\star}^{-2/3} \left(     \frac{\sin(\theta)+e\sin(\phi)}{ \sqrt{1-e^{2}} }  \right)  
\label{eq445}
\end{equation}
\noindent
as the star's velocity amplitude, and $\rho$ given by
\begin{equation}
\rho =   \frac{e^{hc/k\lambda T_{\mathrm{eff}}} \left(  3-hc/k\lambda T_{\mathrm{eff}}   \right)-3}{e^{hc/k\lambda T_{\mathrm{eff}}}-1} .
\label{eq446}
\end{equation}

\noindent
Explanations of Eqs. (\ref{eq444} - \ref{eq446}) and of the underlying phenomenon are given by \citet{2003ApJ...588L.117L}. If this Doppler shift 
can be measured in addition to the variations induced by the stellar ellipsoidal deformation, then indeed the planetary mass can be constrained 
independently from the stellar mass. \par
In the next section, we will firstly explore constraints on accuracy raised by our method and then, as test cases, we will derive the mass of the 
known transiting exoplanet HAT-P-7, based on Kepler data. Finally, we predict the value of our method for the upcoming ESA mission PLATO.

\subsubsection{Thermal emission}

\begin{figure}
\centering
 \includegraphics[width=9cm]{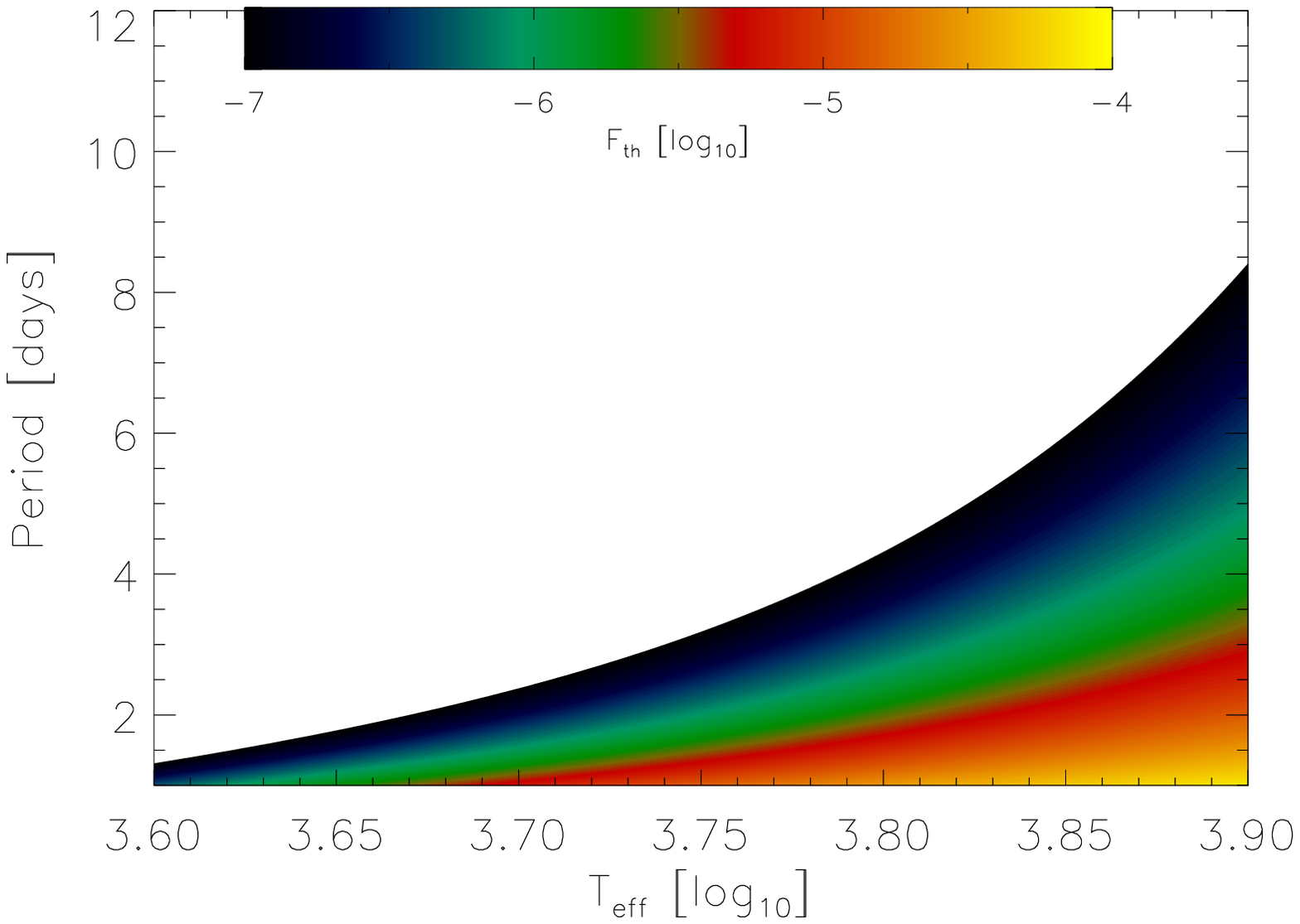}
 \includegraphics[width=9cm]{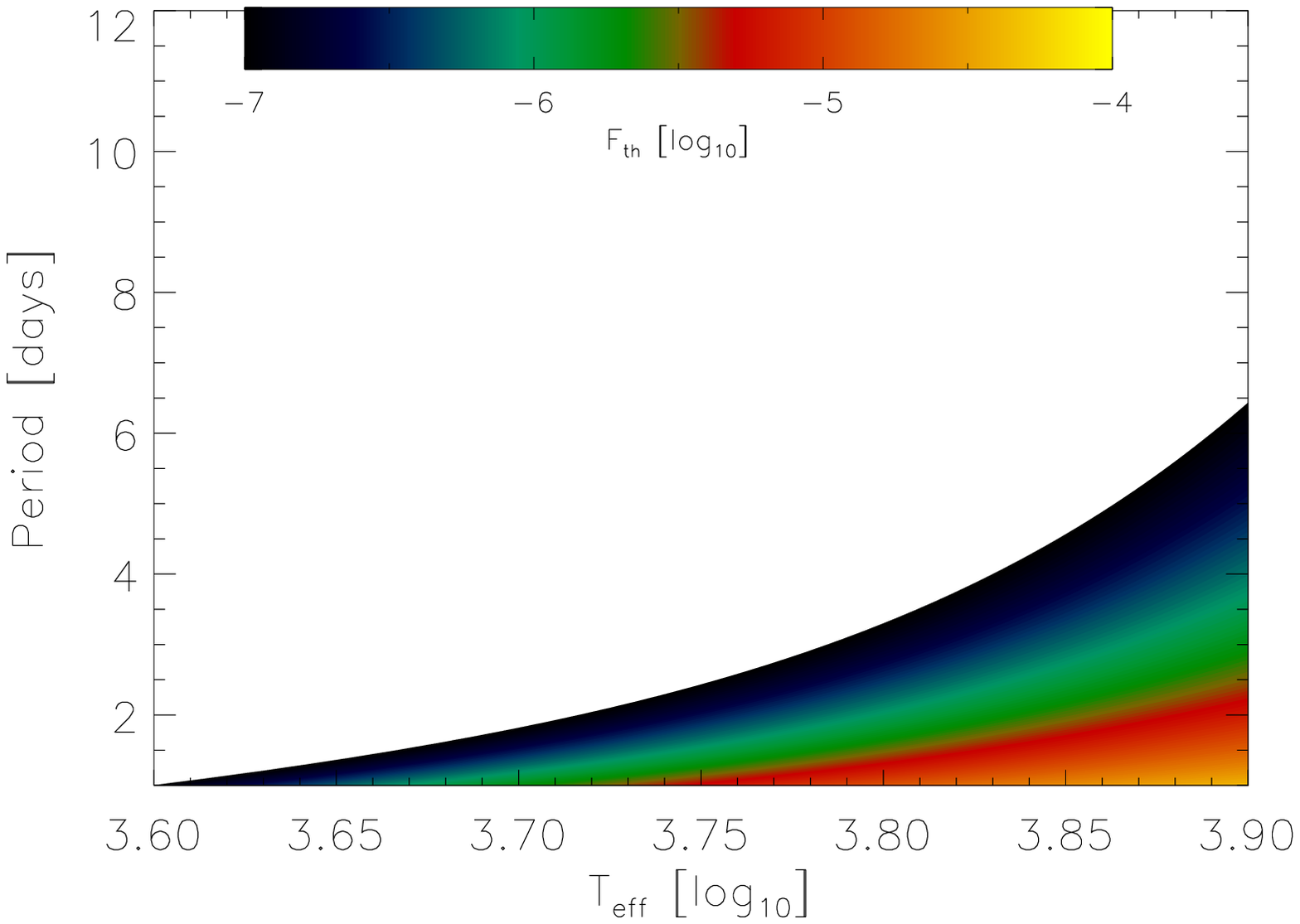} \\
  \includegraphics[width=9cm]{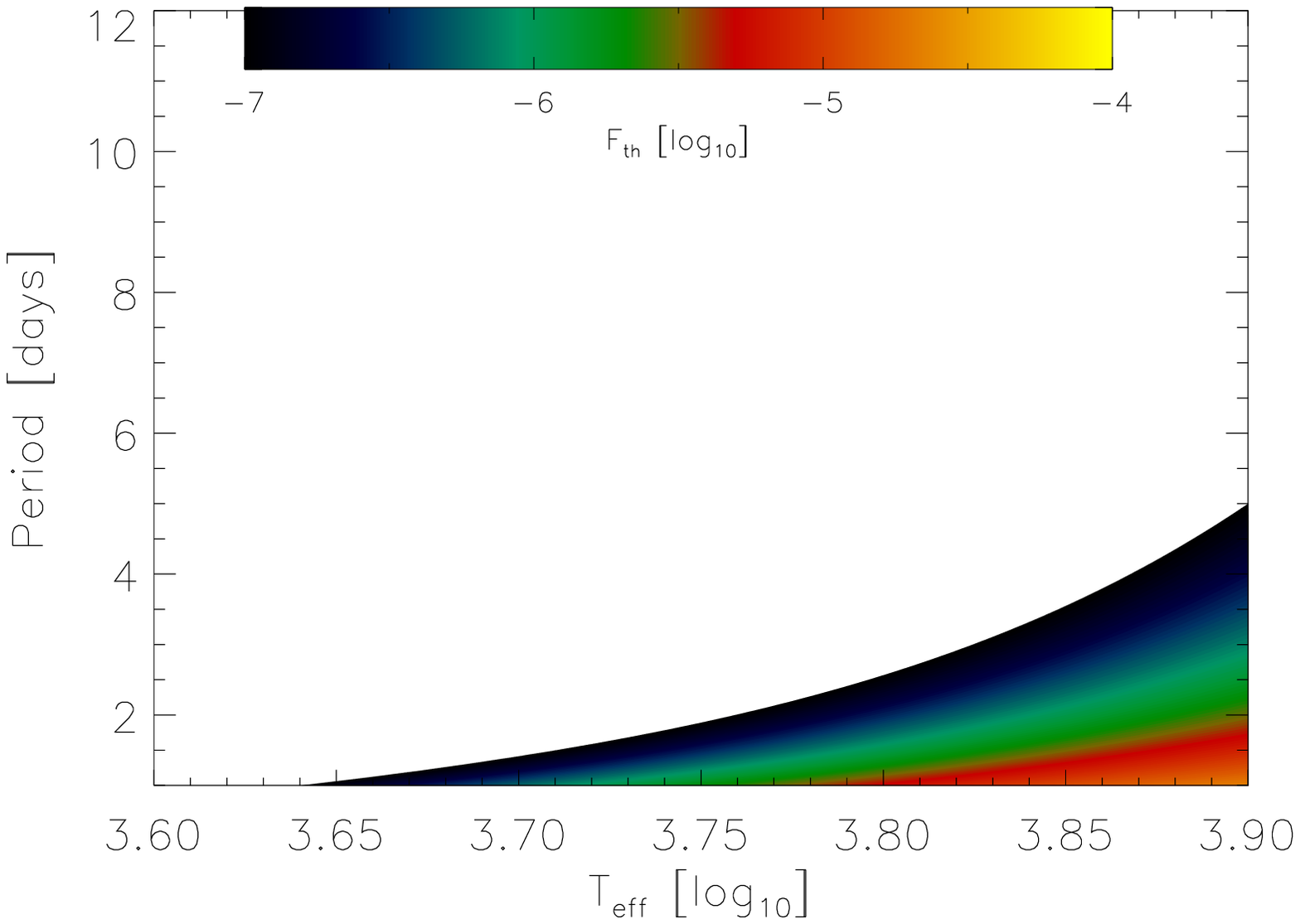} \\
 
 \caption{Thermal emission flux maps for three different bolometric albedo values (top : $\alpha_\mathrm{bol}=0.0$, middle $\alpha_\mathrm{bol}=0.3$, 
bottom: $\alpha_\mathrm{bol}=0.5$). Thermal emission is stronger for short term period planets around hot stars. For all the plots, we assume $R_{P}=1.0R_{J}$.
 The white part of the diagram refers to $f_{th} < 10^{-7}$.}
 \label{thermal}
\end{figure}

The equations for the planetary emitted light are similar to those from \citet{2010arXiv1011.0428C}. Thus 
\begin{equation}
\frac{f_\mathrm{th,d}(z)}{f_\star} = \Phi({z}) \left( \frac{R_{\mathrm{p}}}{R_{\star}} \right)^{2}  \frac{e^{hc/ \lambda k T_{\mathrm{eff}}}-1}{e^{hc/\lambda k T_{\mathrm{d}}}-1}
\label{eq82}
\end{equation}
\noindent
and 
\begin{equation}
\frac{f_\mathrm{th,n}(z)}{f_\star} = {\Big(} 1-\Phi({z}) {\Big)} \left( \frac{R_{\mathrm{p}}}{R_{\star}} \right)^{2}  \frac{e^{hc/ \lambda k T_{\mathrm{eff}}}-1}{e^{hc/\lambda k T_{\mathrm{n}}}-1} ,
\label{eq83}
\end{equation}
\noindent
$T_{\mathrm{eff}}$ is the effective temperature of the star, and $T_{\mathrm{d}}$ and $T_{\mathrm{n}}$ are the 
temperatures on the day and night side of the planet, respectively. Using the equations from \citet{2008ApJS..179..484H} and 
\citet{2010arXiv1011.0428C}, we calculate the mean temperature of the planetary day and night sides:

\begin{equation}
T_\mathrm{d} = T_\mathrm{o} \left( 1-\alpha_\mathrm{bol} \right)^{1/4} \left( \frac{2}{3} - \frac{5}{12}\epsilon  \right)^{1/4}  
\label{eq441}
\end{equation}
\noindent

\begin{equation}
T_\mathrm{n} = T_\mathrm{o}  \left( 1-\alpha_\mathrm{bol} \right)^{1/4} \left( \frac{\epsilon}{4}  \right)^{1/4} ,
\label{eq442}
\end{equation}
\noindent
where $T_{\mathrm{o}}~=~T_{\mathrm{eff}}(R_{\star}/d)^{0.5}$ and $0~\leq~\epsilon~\leq~1)$ is the energy circulation. \par

 \begin{figure}[h]
\centering
 \includegraphics[width=9cm]{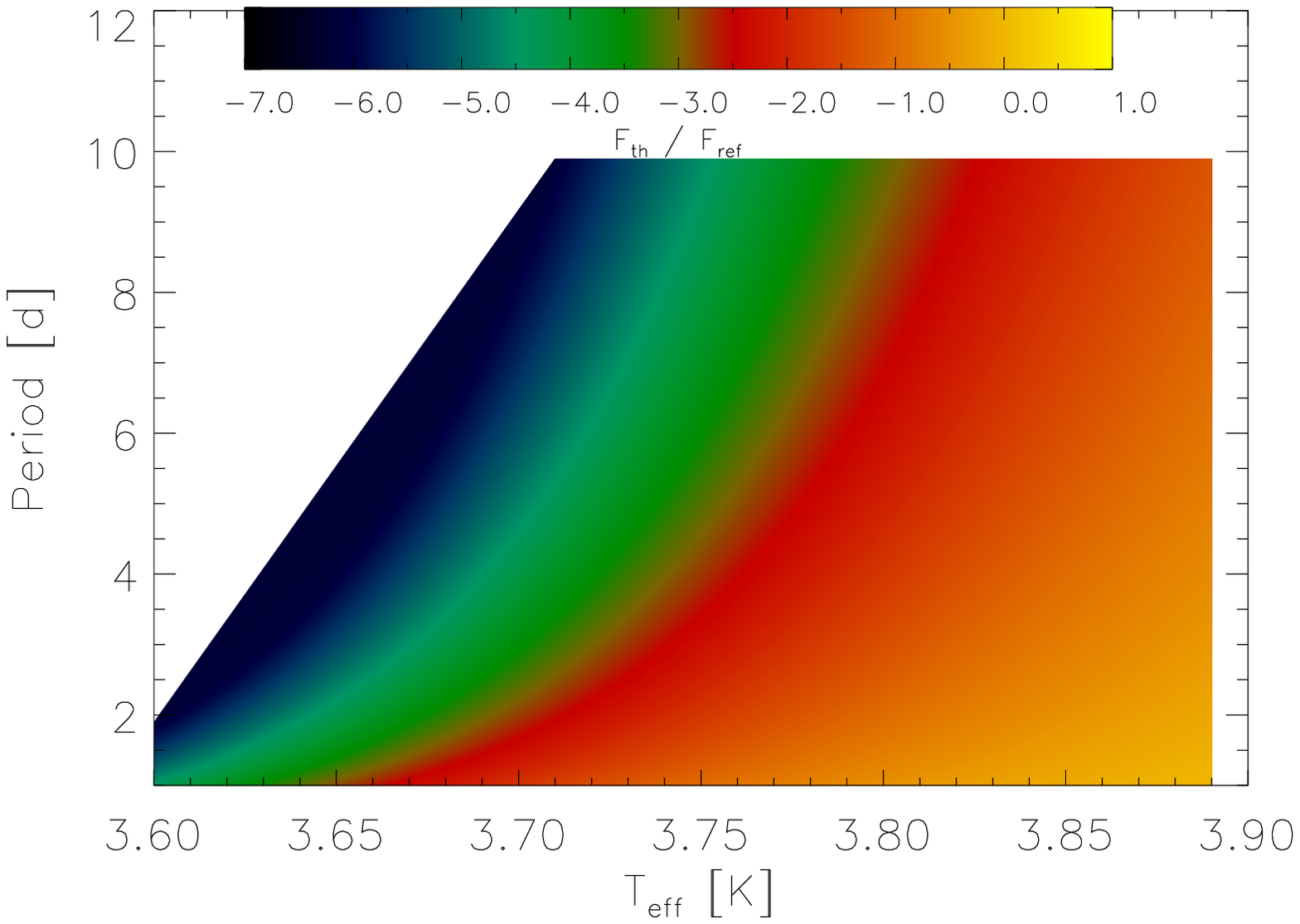} \\
  \includegraphics[width=9cm]{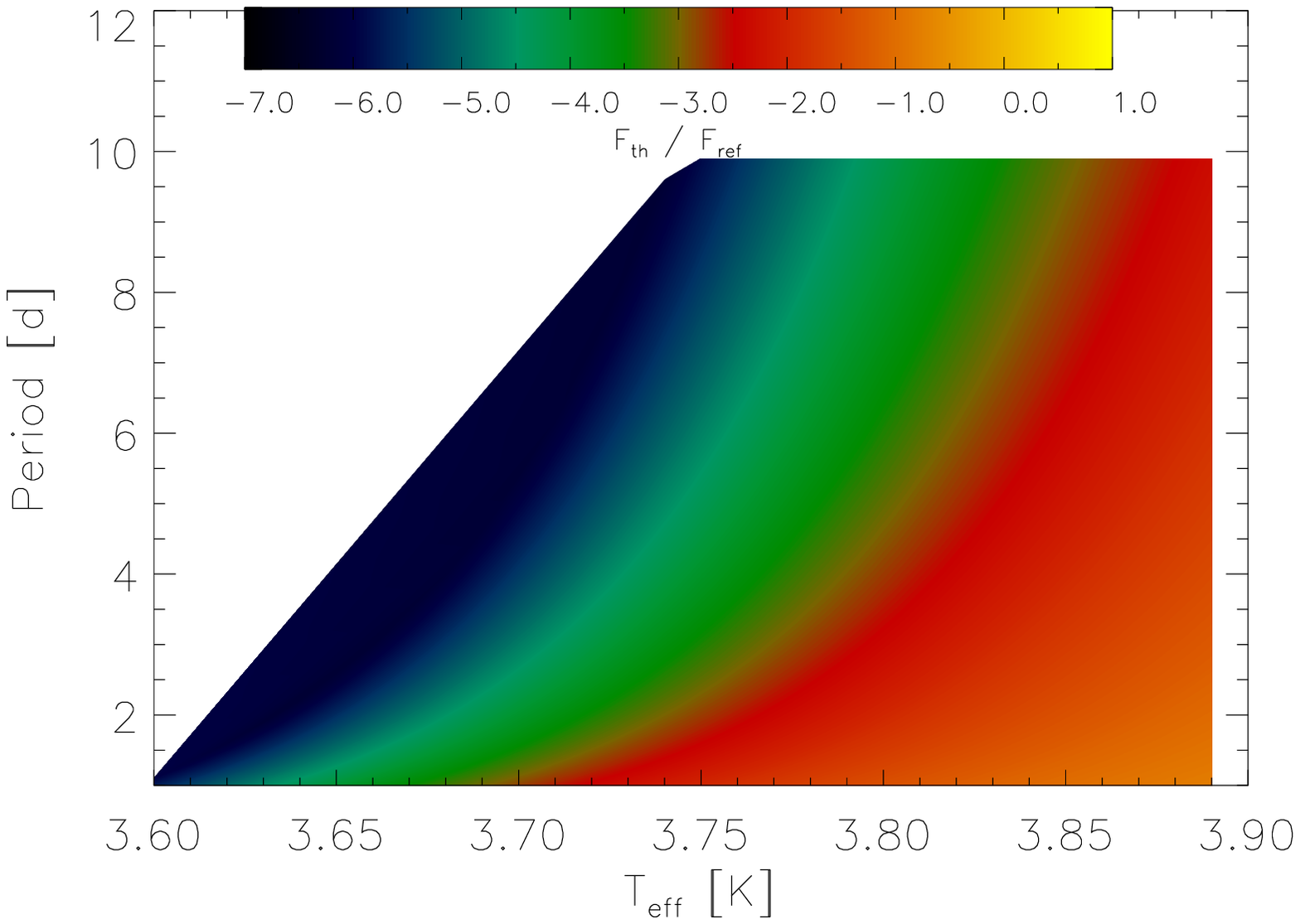} \\
  \includegraphics[width=9cm]{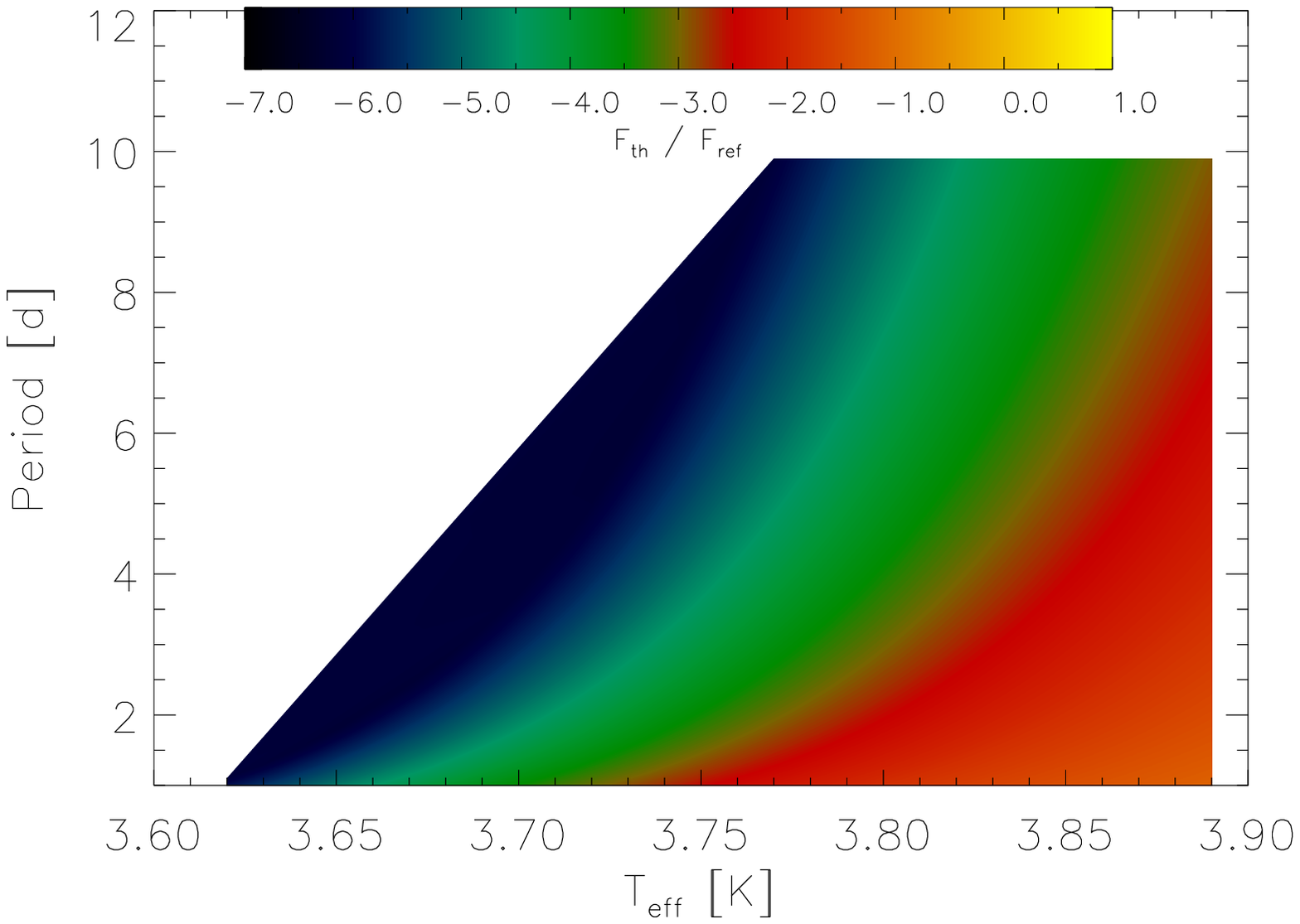} \\
 \caption{Color map of $f_\mathrm{th}$ / $f_\mathrm{ref}$ as a function of $T_{eff}$ and Period for three different albedo cases 
($\alpha_\mathrm{g}$,$\alpha_\mathrm{bol}$=0.05, $\alpha_\mathrm{g}$,$\alpha_\mathrm{bol}$=0.3, $\alpha_\mathrm{g}$,$\alpha_\mathrm{bol}$=0.5 for the top, middle and bottom figure 
respectively). We have fixed the energy circulation to $\epsilon$ = 0.5 and $R_{\mathrm{P}}$=1.0$R_{\mathrm{J}}$. The white part of the diagram
refers to $f_{th}/f_{ref} < 10^{−7}$.}
 \label{fig80} 
\end{figure}
\noindent

The spectrum of the planet will strongly depend on the absence or presence of clouds, the magnitude of 
the greenhouse effect and non-grey opacities. For the most of our analysis we assume that the planet emits as a black body.
Anyhow, the differences between a black body and a non-black body emission are small \citep[a factor of 1.5-3.5, see][]{2008ApJS..179..484H}. 
Assuming main sequence stars and planetary energy 
circulation $\epsilon = 0.0$ (for the maximum thermal emission assuming black bodies), we found that for most hot Jupiters, the thermal 
emission from the planet is very weak in optical wavelengths. For our analysis we assume a blackbody, observed through the Kepler's 
sensitivity curve, which peaks at 575nm and covers 420 to 900nm (Kepler Handbook). Figure \ref{thermal} shows the star's effective temperature $T_{\mathrm{eff}}$ 
versus the period of the planet for three different bolometric albedo $(\alpha_\mathrm{bol})$ values. Colors refer to the expected stellar relative flux\par
The thermal contribution only becomes comparable to, or even larger than, the reflected light component for short period, low albedo planets
 around hot stars. Distinguishing the relative contributions from a LC measured in a single passband, without any knowledge of the planet's albedo is hard, if not impossible. Both thermal and reflected components will have similar phase shapes. In Figure \ref{fig80} we compare $f_{th}$ and $f_{ref}$ for three albedo cases ($\alpha_\mathrm{g}$,$\alpha_\mathrm{bol}$=0.05, $\alpha_\mathrm{g}$,$\alpha_\mathrm{bol}$=0.3, $\alpha_\mathrm{g}$,$\alpha_\mathrm{bol}$=0.5).  \par 
The ellipsoidal variations occur on half the orbital period, and present a rather different phase dependent LC. Even if we cannot accurately 
distinguish between thermal and reflected components, we show in this paper, that they can be jointly distinguished from ellipsoidal variations 
(Fig. \ref{fig68}).  \par

\section{Results}
\label{results}

\subsection{Simulations}

To assess the data accuracy required by our method to yield satisfying constraints on the planetary parameters, we simulate two LCs using the 
equations from above. One of our model planetary systems is a hot Jupiter, analogue to the transiting planet HAT-P-7b \citep{2009MNRAS.tmp.1781P}, 
whereas the other one resembles the transiting Super-Earth CoRoT-7b \citep{2009A&A...506..303Q}. We customize these models in terms of the 
geometric albedo, for which we optimistically apply $\alpha_\mathrm{g}=0.3$ in both cases \citep{2000ApJ...538..885S}. Though observations of 
CoRoT-7b are reconcilable with $e=0$, we chose $e= 0.05$ to simulate enhanced phase variations in the LC. After all, we are not heading for a reconstruction 
of these systems but we want to estimate how accurate comparable systems could be parametrized and, as an example, if a putative small eccentricity 
of CoRoT-7b could be determined. This is particularly important given the significant activity of CoRoT-7b. 

\begin{table}
\centering
\caption{Physical parameters of our two models. For the Super-Earth planet we assume no thermal emission $(\alpha_\mathrm{bol} = 1.0)$}
\label{tab1}

\begin{tabular}{ccc}

   \hline

   \hline

   \hline \hline

 Stellar parameter & hot Jupiter & Super-Earth\\

 \hline

 \noalign{\smallskip}

 $R_\star$                      &  $1.64\,R_\odot$   &  $0.87\,R_\odot$  \\
 $M_\star$                      &  $1.36\,M_\odot$   &  $0.93\,M_\odot$  \\
 $T_{\mathrm{eff}}$  &          6290\,K      &         5275\,K       \\
$\beta$                            &              0.53                &          0.61               \\
$\epsilon $                      &          0.75                &              \\
 \\
\end{tabular}

\begin{tabular}{ccc}
 Planetary parameter & hot Jupiter & Super-Earth\\

 \hline

 \noalign{\smallskip}

 $R_\mathrm{p}$         &   $1.16\,R_\mathrm{J}$   &   $0.15\,R_\mathrm{J}$     \\
 $M_\mathrm{p}$        &    $8.75\,M_\mathrm{J}$  &   $0.0151\,M_\mathrm{J}$  \\
 $\alpha_\mathrm{g}$  &               0.30              &         0.30                        \\
  $\alpha_\mathrm{bol}$  &               0.65              &         1.0                        \\
 $u_1$                       &               0.34              &         0.20                       \\ 
 $u_2$                       &               0.35              &         0.57                       \\  
 \\
\end{tabular}

\begin{tabular}{ccc}
 Orbital parameter &   hot Jupiter   &   Super-Earth  \\

 \hline

 \noalign{\smallskip}

 $P$            &    5.63347\,d        &      0.85360\,d       \\
 $i$             &    $86.72^\circ$    &       $80.10^\circ$   \\
 $e$            &      0.52              &       0.05               \\
 $\omega$    &  $72^\circ$         &      $5^\circ$         \\

           \hline
\end{tabular}
\end{table}

\begin{figure*}
\centering
 \includegraphics[width=9cm]{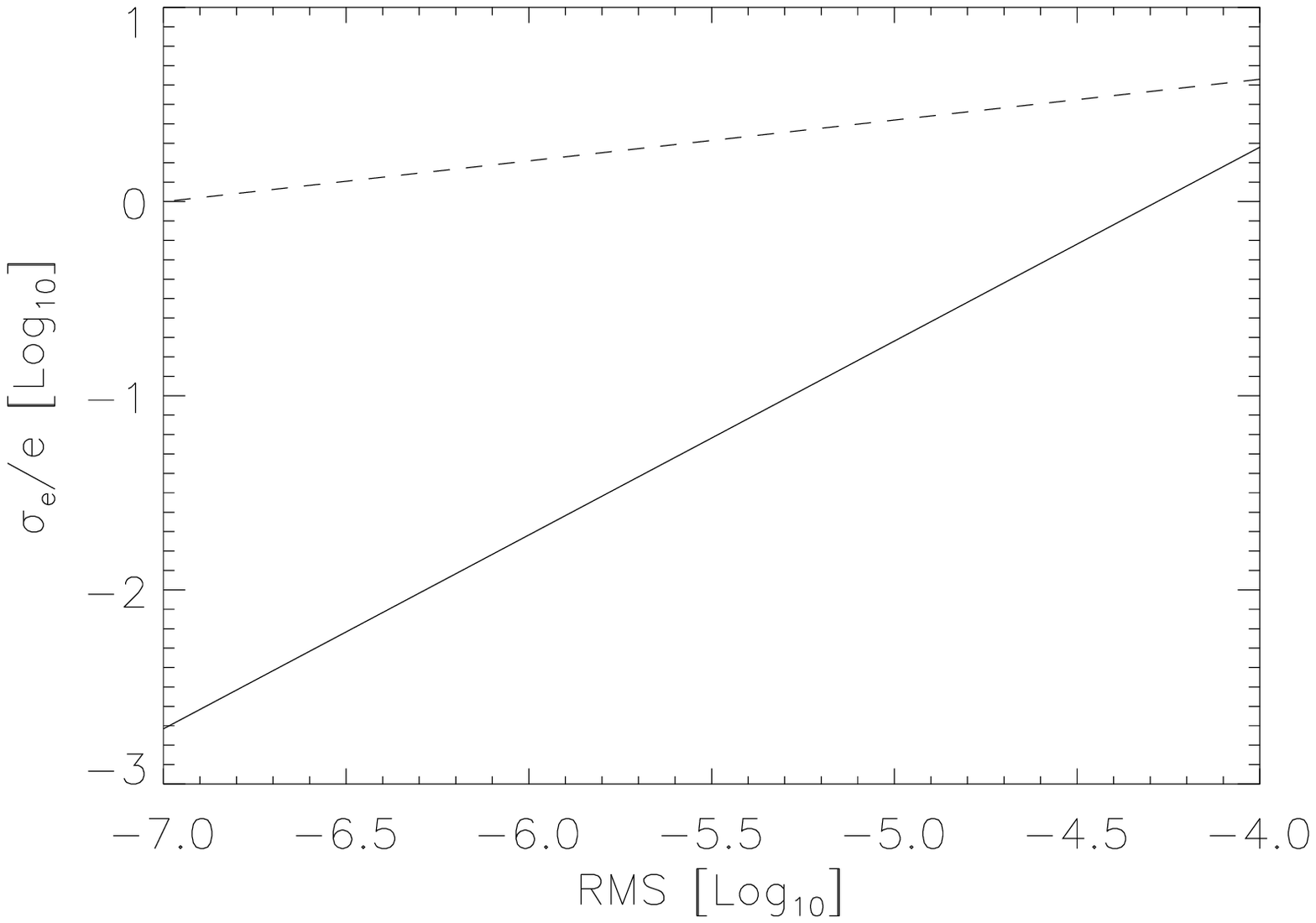}
 \includegraphics[width=9cm]{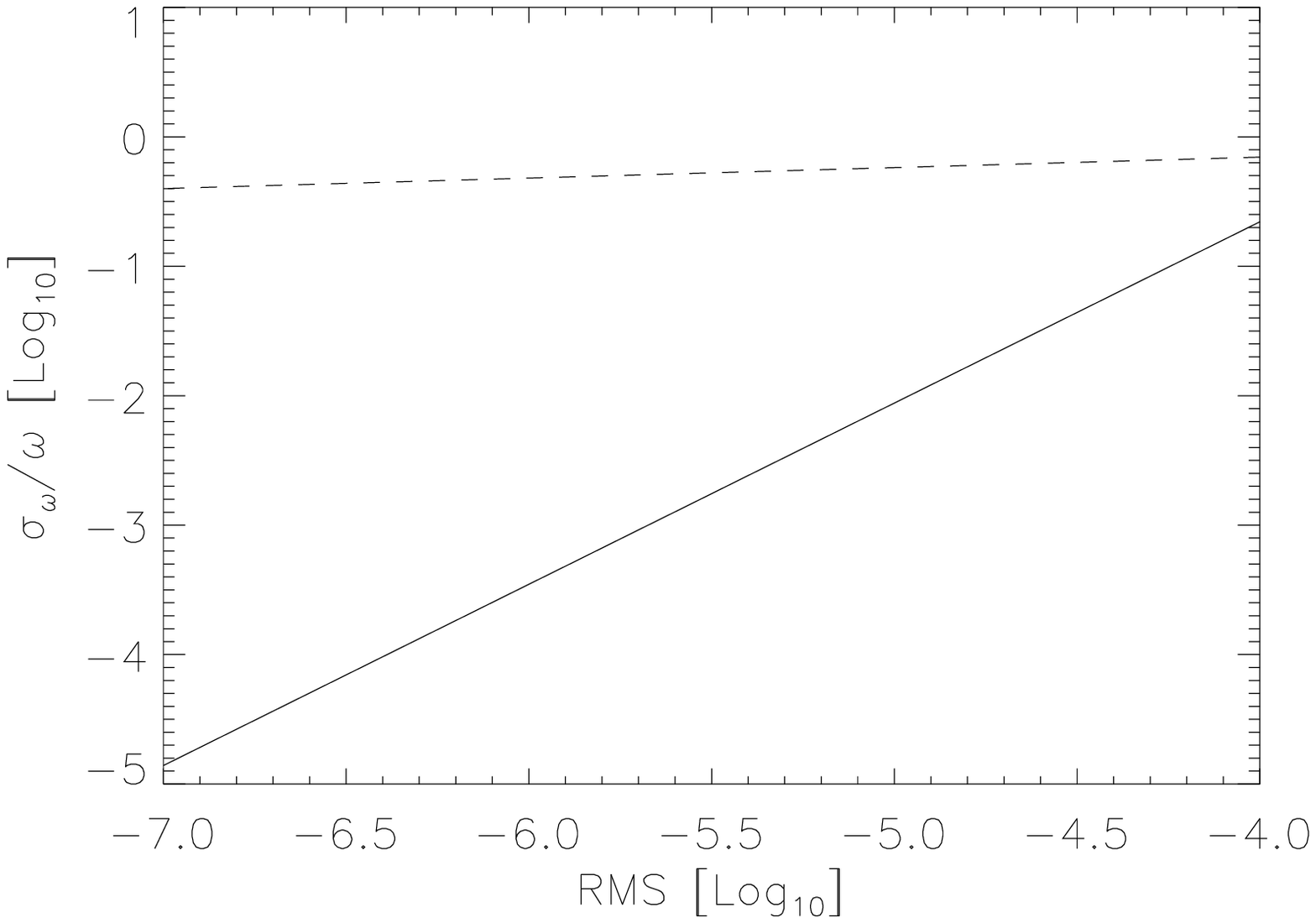} \\
  \includegraphics[width=9cm]{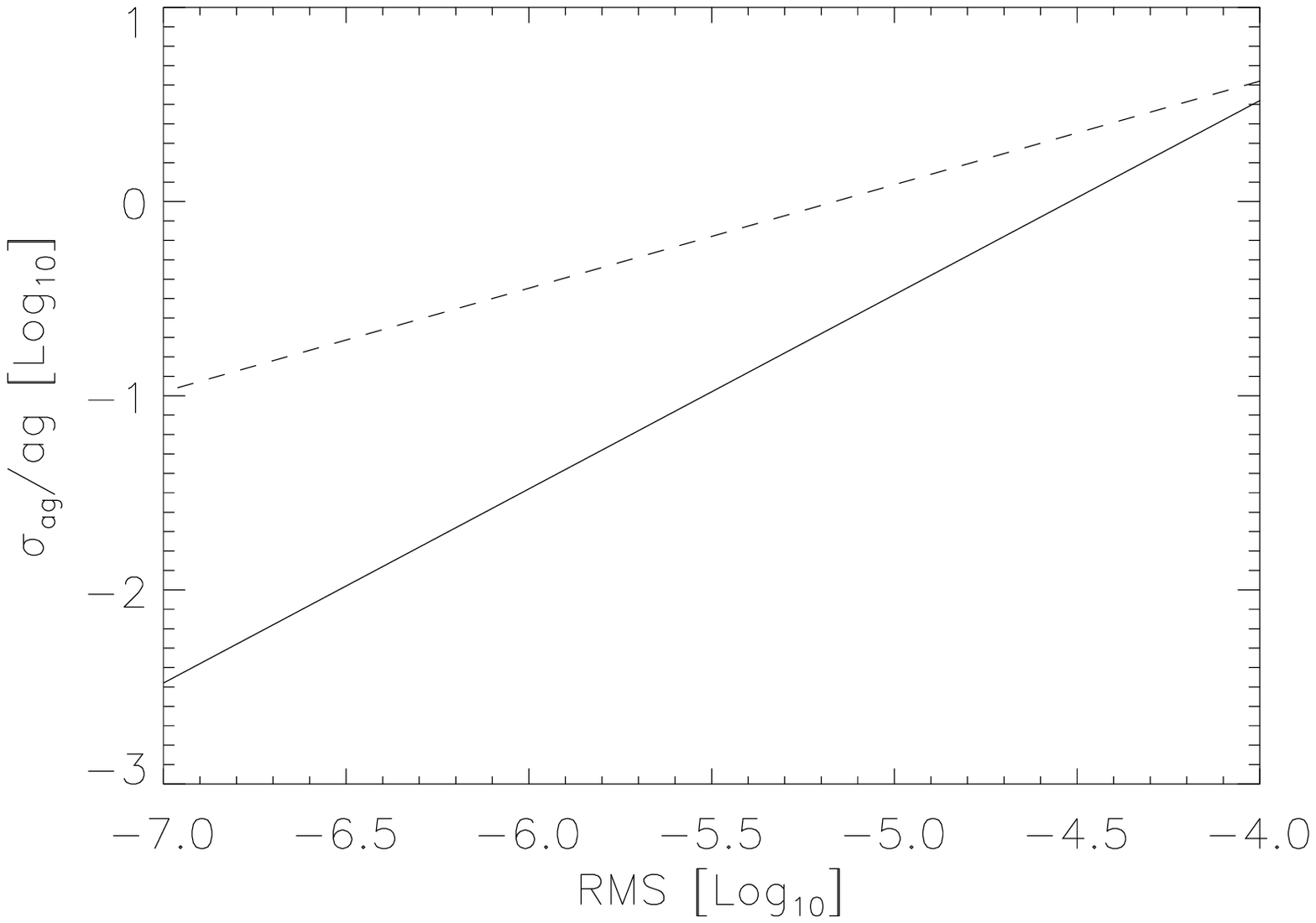} 
 \includegraphics[width=9cm]{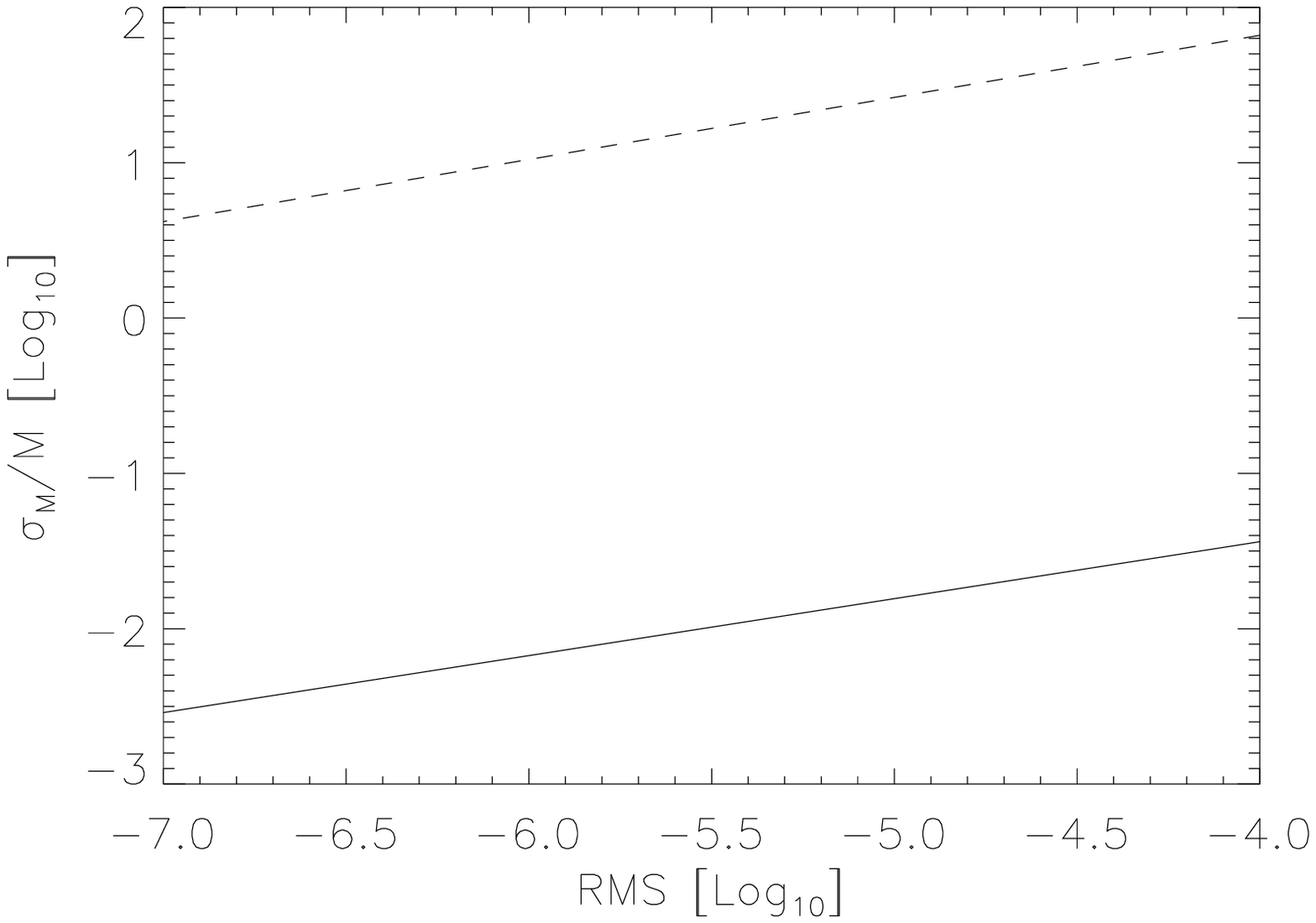}
 
 \caption{Relative errors in $e$, $\omega$, $\alpha_\mathrm{g}$ and  $M_{\mathrm{P}}$ as functions of data accuracy. The solid line 
denotes the HAT-P-2b twin while the dashed line labels the CoRoT-7b analog.}
 \label{fig3}
\end{figure*}

To these models, we add increasingly more noise simulating data accuracies between $10^{-7}$ and $10^{-4}$. The phase effect in the LCs, i.e. 
in $f_3$, is significant only for accuracies $\lesssim 10^{-4}$, which is why this phenomenon has not been detected in the LCs of CoRoT 
\citep{2004ESASP.554..281C}. We then fit the noiseless model from Sect. \ref{sec:theory} to each of these -- more or less -- noisy LCs 
and use 1000 Monte Carlo simulations to calculate the standard deviations for each parameter in each fit using a bootstrap method explained 
by \cite{2008A&A...487L...5A}. In Fig. \ref{fig3} we show the relative errors resulting from these fits for the planetary geometrical albedo 
$\alpha_\mathrm{g}$ ($\sigma_{\alpha_\mathrm{g}} / \alpha_\mathrm{g}$), eccentricity ($\sigma_{e}/e$), orientation of periastron ($\sigma_{\omega}/ \omega$) 
and the mass ratio ($\sigma_{M}/M$) as a function of the root mean square (RMS) of the data. With an accuracy provided by Kepler of $10^{-4}$ 
\citep{2010ApJ...713L..79K}, the eccentricity of a CoRoT-7b-like Super-Earth could merely be determined with a 
unsatisfactory relative error higher than 1.0 which has no physical use. For a hot Jupiter similar to HAT-P-2b, however, the relative 
error in $e$ is about 0.001 in the best case (hot Jupiter - RMS $10^{-7}$). The orientation of periastron for the Super-Earth could be constrained 
to approximately $\pm 4^\circ$ in the worse case (RMS $10^{-4}$) while for the hot 
Jupiter the accuracy is as low as $\pm 0.001^\circ$ in the best case (RMS $10^{-7}$). Restrictions of the geometrical albedo $\alpha_\mathrm{g}$ are 
$\pm0.001$ in the best case (hot Jupiter - RMS $10^{-7}$) to $\pm$ 1.1 in the worst case (Earth like - RMS $10^{-4}$) with no physical meaning. The 
respective accuracies in planet mass $M_{P}$ are $\pm0.028$ $M_{J}$ for hot Jupiters (best case) and $\pm0.95$ $M_{J}$for Super-Earths 
(worst case).

\subsection{Relative Contributions}
\begin{figure}
\centering
\includegraphics[width=9cm]{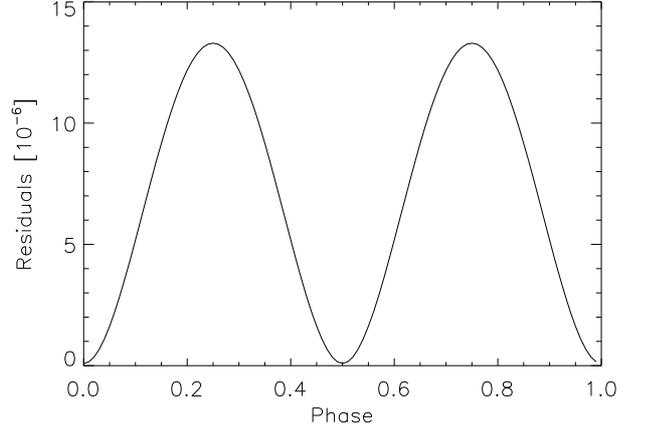}
 \caption{Residuals flux between Lambert \& geometrical sphere. The residuals curve is identical to the ellipsoidal variations curve. By selecting the wrong phase function we over/under estimate the planetary mass.}
\label{fig79}
\end{figure}

Different phase functions (eqs. (\ref{eq5}) \& (\ref{eq4})) lead to  different peaks in the thermal emission and reflected LC. The 
ratio between thermal emission and reflected light over ellipsoidal variations becomes smaller (for the Lambert sphere) or larger (for the 
geometrical sphere), which is why different phase functions lead to different mass values. The mass difference $(\Delta M_\mathrm{P})$ between 
the two cases is a constant (36.4\%). For example, if we use a wrong sphere, the mass we measure for HAT-P-7b exoplanet is 
$\Delta M_\mathrm{P} \sim 0.67M_{J}$ smaller (geometrical) or larger (Lambert) than the true mass. Seeking out the most plausible phase function, we fit two models, both differing only in the assumption of either the Lambertian or the geometrical sphere, to the data. For most systems, if the 
components $f_{ref}$, $f_{th}$ \& $f_{ell}$ are very weak, it is impossible to distinguish between the two phase functions. In this case we select 
the Lambert sphere in order to calculate the upper mass limit of the planet. In Fig. \ref{fig79} we show the residuals between the two different phase functions. Notice that the shape and the phase (x-axis) of residuals is identical to ellipsoidal variations curve.  \par

For most hot Jupiter systems thermal emission in optical wavelengths is much weaker than variations due to the varying projected shape of the star. 
For orbital periods of 1 and 5 days of a G0V star, ellipsoidal variations are 3 and 10 times larger than thermal emission, respectively, while 
$f_\mathrm{th}/f_\star~\sim~10^{-5}$ and $10^{-9}$ (Fig. \ref{thermal} \& \ref{fig80}). As the star becomes hotter (F or A type), or the period 
shorter ($<5$ days), the thermal component become to be more significant. \par
Thermal emission, stellar ellipsoidal deformations, and Doppler shift changes are very weak phenomena (fainter than $10^{-5}$), which require very 
high-accuracy observations to be detected. So far, it has been impossible to detect these effects from the ground but based on data from the 
recent space-based mission Kepler, our system of equations offers a new gate for exoplanet mass determinations. Neither RV measurements nor transit 
timing variations are necessary for our procedure.\par
Our parametrization of reflected and thermal planetary emission are nearly identical, so our model is by definition 
unable to distinguish between these two effects. Both equations \ref{eq3} and \ref{eq82} are only function of the 
orbital phase $(\Phi_{L,C})$ and both components (thermal and reflected) provide exactly the same LC. We are not able to 
distinguish between the two components by fitting our model in a hot Jupiter system, which the planet is tidal locked by its host star. If the 
planetary rotation and orbital period are different (for a case of a high eccentric orbit and the simple assumption that there are no 
winds), then reflected light and thermal emission LCs will show different shapes.


\subsection{Application to Kepler data: The Case of HAT-P-7b }
Kepler is a space-based mission, launched on Mar. 7, 2009, with the primary goal to detect Earth-sized and smaller planets in the habitable 
zone of solar-like stars. For a $m_{V}=12$ star, the photometric precision is $\log_{10}(\mathrm{RMS}) \sim$ $-4.35$ for 6 hr combined 
differential photometry \citep{2011ApJ...736...19B}. We choose the well-studied transiting planet HAT-P-7 in the Kepler field, with mass and 
system parameters constrained by both photometric and RV observations, to test our photometry-only method. We use all the short exposure data 
(1 minute exposures) from Kepler's public archive. Our sample includes 67 transits and the phase folded LC was made by $\sim 180,000$ epochs 
binned by a factor of $\sim 50$. \par  
\citet{2010ApJ...713L.145W} already found ellipsoidal variations in HAT-P-7. Using the established physical parameters of the system, we calculate that 
the effect of $(f_{\mathrm{ell}}~\sim~3.1\cdot10^{-5})$ is $\sim3.56$ times more pronounced than the Doppler shift variations 
$(f_{\mathrm{dop}}~\sim~8.7\cdot 10^{-6})$. Nevertheless, we include both effects in our model.\par
\begin{figure}
\centering 
 \includegraphics[width=9cm]{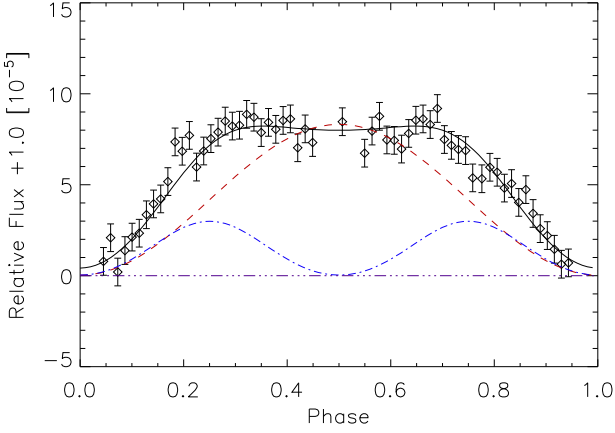}
 \caption{LC of HAT-P-7. The solid line shows the best fit solution of our synthetic model, i.e. including stellar ellipsoidal distortions and Doppler effect 
variations. The dashed red line refers to the reflected light component, the blue dotted line refers to the ellipsoidal variations and the purple, 
almost consant dashed line refers to the thermal emission. The Lambert sphere assumption produced the best fit.} 
 \label{fig68}
\end{figure}

We assume that the stellar radius and effective temperature are known $(T_{\mathrm{eff}}$ = 6350$K)$. We fix the ratio of stellar and 
planetary radius $R_{\mathrm{p}}/R_{\star}$ = 0.0778, the inclination $i = 83.1^\circ$ and the relative semi-major axis $a/R_{\star}$ = 8.22 
\citep{2010ApJ...713L.145W}. For our first fit, we use Eq. \ref{eq8}, but we exclude $f_{ell}$ and $f_{dop}$. We only fit $f_{ref}$ plus $f_{th}$
in order to check if only these two components could successfully explain the data $(\chi_{\mathrm{ref+th}}^{2})$. Our second fit includes all four 
components from Eq. \ref{eq8} $(\chi_{\mathrm{all}}^{2})$. The $\chi^2$ ratio between the reflected light plus the thermal emission 
model and the model which includes all four components is $\chi_{\mathrm{ref+th}}^{2} / \chi_{\mathrm{all}}^{2} = 3.04$. Using an F-test, we found 
that the complete model explains the data better than the simple model of the reflected plus thermal emission. There is a $0.01\%$ chance that the 
more complex model do not produce a better fit to the data (F-Ratio:116.28, P-Value: 0.0001). In Fig. \ref{fig68} we show the observed LC of 
HAT-P-7  with our best fit as well as a comparison to a model, which includes only reflected stellar light from the planet, and our enhanced model. 
In Table \ref{tab2} we list the output parameters of the fit. Because $f_{th}$ is very weak ($10^{-6}$ - $10^{-7}$, see Fig. \ref{thermal} for the
black body case), $\alpha_\mathrm{bol}$, $T_{d}$ and $T_{n}$ values can not be calculated accurately (very large errors). \cite{2010ApJ...713L.145W}, 
used the secondary eclipse depth of HAT-P-7b, in order to measure the flux from the night side of the planet ($f_{th,n} = 2.2 \cdot 10^{-5}$). 
From Fig. \ref{thermal} (bottom diagram) is clear that the planet is much brighter and much hotter than expected, and other heating mechanisms 
might exist. We can not disentangle $f_{ref}$ from $f_{th}$, however this does not affect our main result. \par
We find the ratio of the planetary mass over the stellar mass to be $M_{\mathrm{p}}/M_{\star}$ = 1.27$M_{\mathrm{J}}/M_{\odot}$, compared to 
$M_{\mathrm{p}}/M_{\star}$ =1.20 $M_{\mathrm{J}}/M_{\odot}$ deduced by \citet{2010ApJ...713L.145W}. Because the $f_{ell}$ component
is uncorrelated with $f_{ref}$ and/or $f_{th}$, the accuracy of the mass ratio $(M_{\mathrm{P}}/M_{\star})$ can be calculated with high accuracy. 
Although our values are not as accurate as those from RV measurements, we can infer the planetary nature of the companion.

\noindent

\begin{table}
\centering
\caption{HAT-P-7b system parameters.}
\label{tab2}

\begin{tabular}{ccc}

   \hline

   \hline

   \hline \hline

 HAT-P-7 parameters & Our model & Other studies\\

 \noalign{\smallskip}
$M_{\mathrm{p}}/M_{\star}$                        &  1.27 $\pm$ 0.10 $M_{\mathrm{J}} / \,M_\odot$  &  1.20 $\pm$ 0.05  $M_{J} / \,M_\odot$ $^{\alpha}$   \\
 $e$                        &  0.0 $\pm$ 0.1    &  0.003 $\pm$ 0.012 $^{\beta}$ \\
 $\phi$                      &  0.0 $\pm$ 1.0 $^{o}$   &  0.0 $^{\beta}$ \\ 
  $\alpha_\mathrm{g}$                        &  0.21 $\pm$ 0.08   &  0.18  $^{\alpha}$ \\
  \\
 \hline
 \hline
 $^{\alpha}$ \citep{2010ApJ...713L.145W} & & \\
  $^{\beta}$ \citep{2009IAUS..253..428P} & &\\
\hline
\end{tabular}

\end{table}

\subsection{PLATO mission}
PLATO is a prospective ESA mission aiming at the characterization of transiting exoplanets. The mission will use 40 small telescopes 
covering $42~\times~42~\mathrm{deg}^2$ in total with a supposed overall accuracy of $\sim~2.8~\cdot~10^{-5}$ \citep{2010Ap&SS.328..319C}. This 
will satisfy the needs of our procedure. In the following, we demonstrate the capabilities of our photometry-only method in parameterizing 
hypothetical exoplanet systems.\par
\begin{figure}[h]
\centering
 \includegraphics[width=9cm]{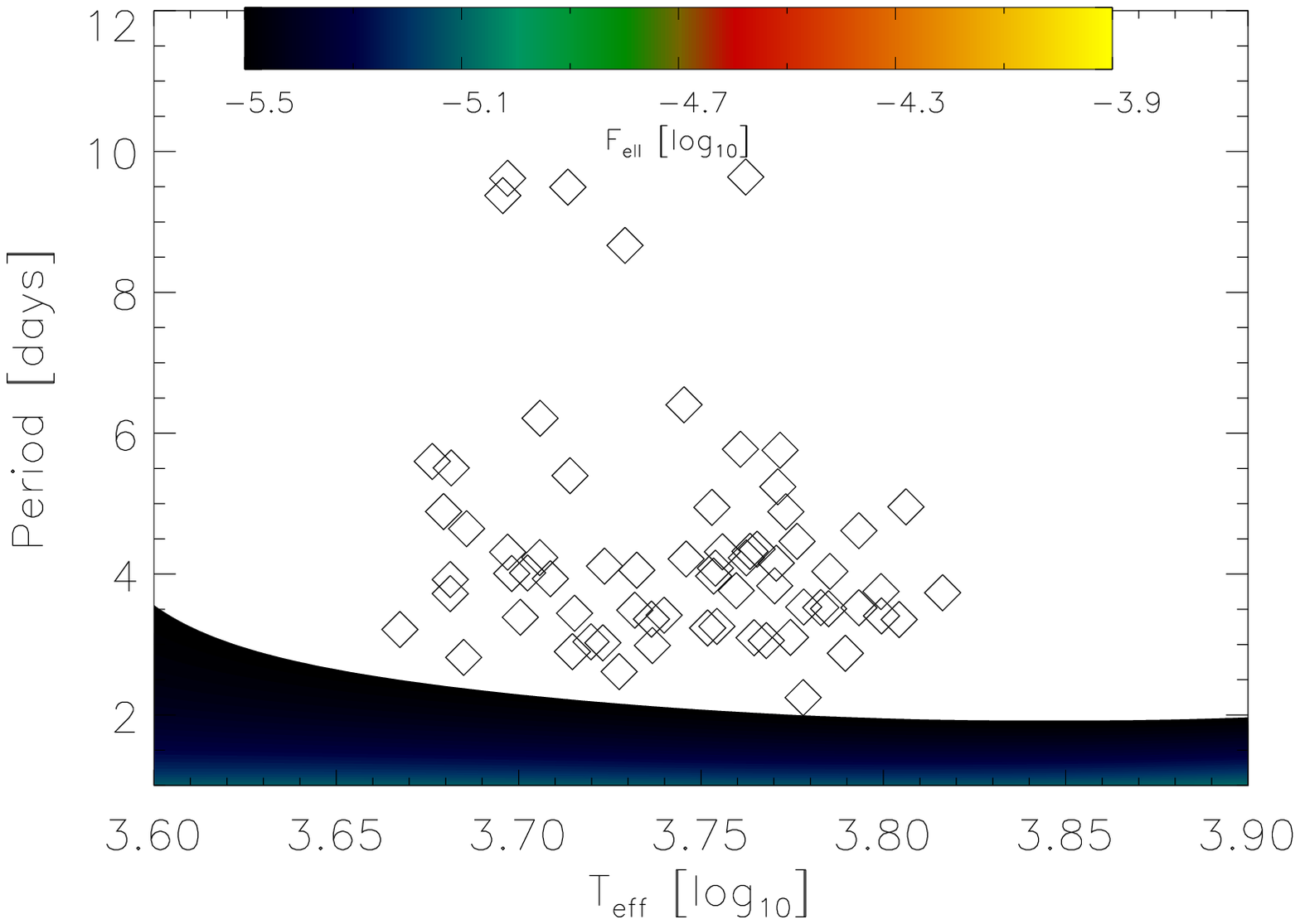}
  \includegraphics[width=9cm]{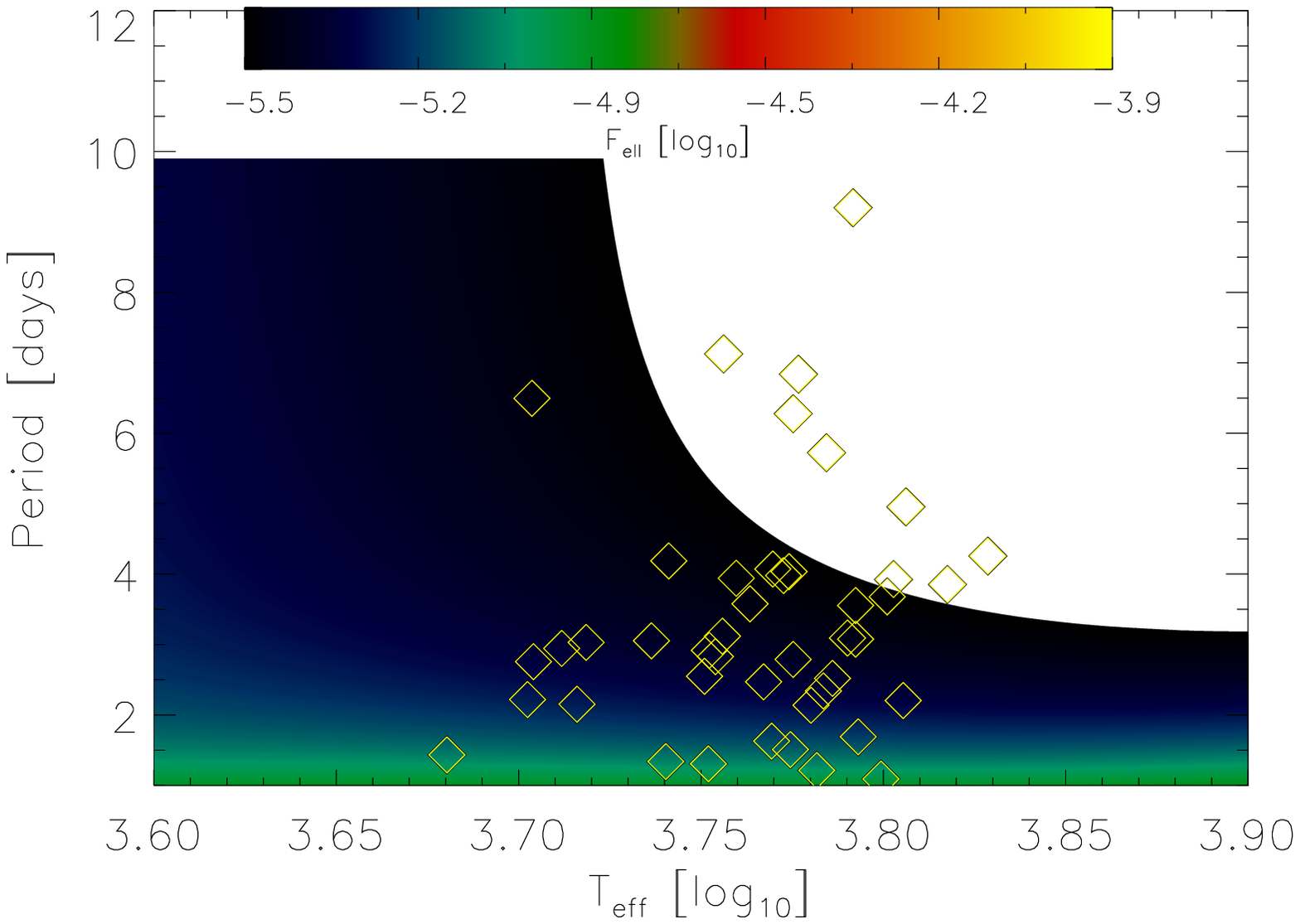}
    \includegraphics[width=9cm]{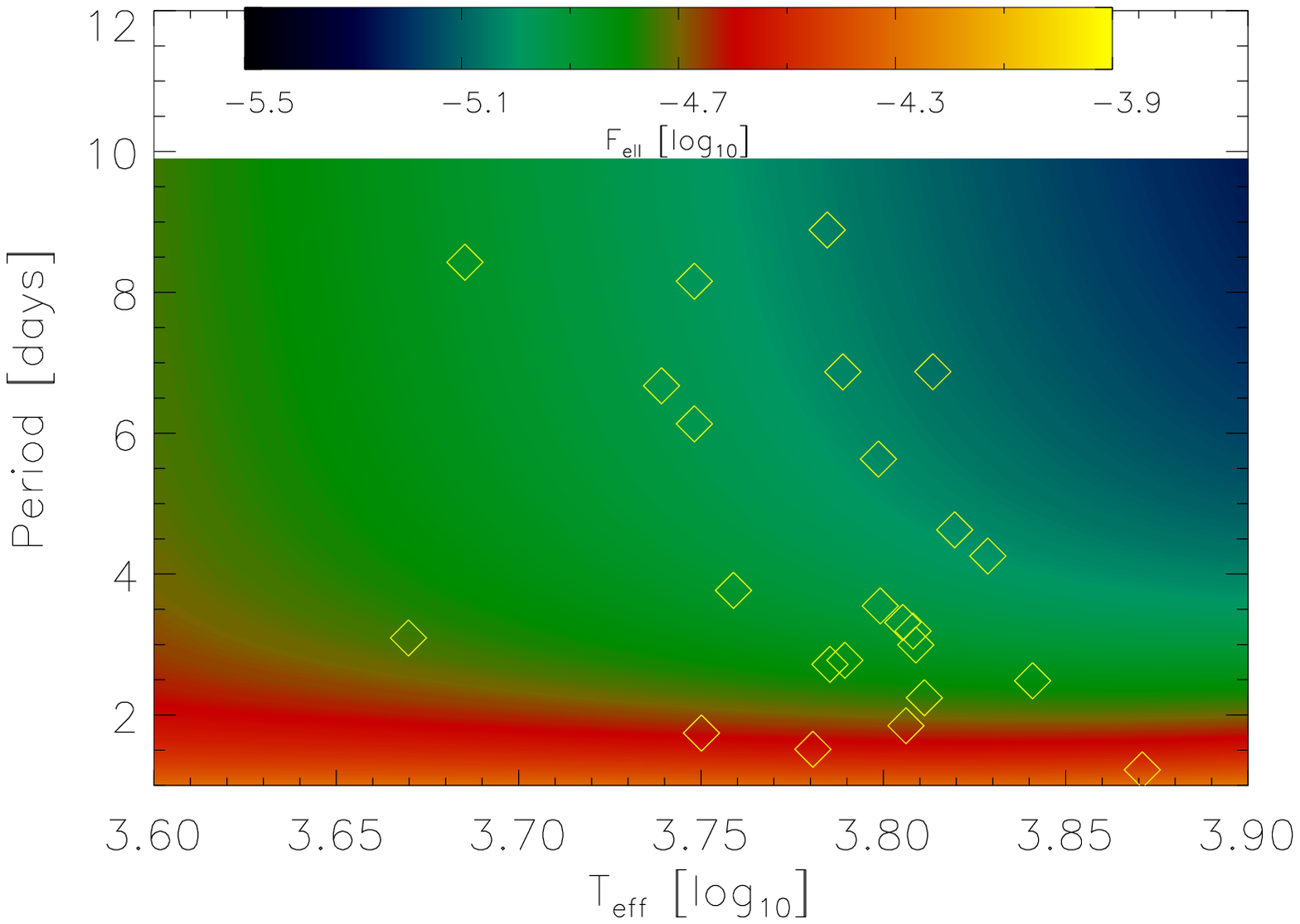}

 \caption{Combination of stellar ellipsoidal distortion and Doppler shift effects in simulated PLATO observations. The panels show the amplitude of 
flux variations $\Delta F$ in the LCs for three different planetary masses: $M_\mathrm{P}~=~0.5\,M_{\mathrm{J}}$ (top), $1\,M_{\mathrm{J}}$ (middle), and 
$5\,M_{\mathrm{J}}$ (bottom) around a range of main-sequence stars. Diamond symbols represent known transiting planets with similar masses. 
The white part of the diagram refers to $f_{ell} < 10^{-5.5}$.}
 \label{fig70}
\end{figure}

We investigate the detectability of $f_\mathrm{ell}$ and $f_\mathrm{dop}$ in a range of different systems, assuming that our target stars 
belong to the main sequence. For each combination we calculate the amplitude of the flux variations $\Delta F$ 
in the LC. In Fig. \ref{fig70} we present three plots, each belonging to a different planetary mass: $M_{\mathrm{p}}~=~0.5\,M_{\mathrm{J}}$ 
(top), $1\,M_{\mathrm{J}}$ (middle), and $5\,M_{\mathrm{J}}$ (bottom). For comparison we indicate some known transiting systems with similar 
masses. In the upper panel $\Delta F$ is dominated by $f_\mathrm{ell}$. Here, for the sample of known systems $\Delta F~>~10^{-5}$, which would be 
challenging to be detected. Nevertheless, with PLATO one will be able to detect $f_\mathrm{ell}$ for planets with $P~<~3\,\mathrm{d}$ around stars 
with spectral type K0 or later (blue zone in the top panel of Fig. \ref{fig70}). In the middle diagram $(M_{\mathrm{p}} = 1M_{\mathrm{J}})$ both 
$f_\mathrm{ell}$ and $f_\mathrm{dop}$ affect the LC and $\Delta F~\sim~5\,\cdot\,10^{-5}$ for most of the known systems -- detectable with 
PLATO. In the bottom panel, where the model planet is most massive, the LCs are mostly affected by the Doppler shift variations rather than by the 
ellipsoidal geometry of the star. The phenomenon is fairly detectable for the most cases of the known planets while $\Delta F~\sim~12 \cdot~10^{-5}$. 

\noindent

\section{Conclusions}
\label{conclusions}

The mathematical tools presented in this article can be used for a complete parametrization of transiting exoplanet systems on the basis of 
high-accuracy LCs only. In our model, for a specific range of systems, RV measurements are not necessary to constrain the mass of the planet 
($M_\mathrm{p}$), orbital eccentricity ($e$), the orientation of periastron ($\omega$), and the geometric albedo of the planet 
($\alpha_\mathrm{g}$). Our model also incorporates the characterization of the ratio of planetary and stellar radius ($R_\mathrm{p}/R_\star$), 
orbital period ($P$), and the orbital inclination ($i$). In order for this method to be fully applicable, the planet must be massive enough and 
orbit its host star close enough as to distort the stellar structure significantly.\par
With the current Kepler and potential future missions such as PLATO, we are able to measure the mass of the hot Jupiters, where the orbital periods 
must be $\leq~1.5$, $\leq~4$, and $\leq~10$ days in order to measure planetary masses of $0.5\,M_\mathrm{J}$, $1\,M_\mathrm{J}$, and 
$5\,M_\mathrm{J}$ respectively. Furthermore, for host stars with spectral types earlier than FV, thermal emission component is 
detectable in the LC (assuming black bodies). This will affect the accuracy and even the detection of ellipsoidal variation. \par
As we show, the Kepler mission provides an accuracy suitable enough for our procedure to be applied for the characterization of extrasolar planets. 
We confirm the planetary natures of the hot Jupiters HAT-P-7 from Kepler data alone. Our method yields masses of 
$M_\mathrm{p} = 1.27^{+0.10}_{-0.45}$ $M_\mathrm{J}~\times~M_\star/M_\odot$ for HAT-P-7b. Our results ($M_\mathrm{P}$, $\alpha_\mathrm{g}$, 
$\alpha_\mathrm{bol}$, $e$, $T_{d}$, $T_{n}$) are in good agreement with \citet{2010ApJ...713L.145W} and we also confirm the ellipsoidal variations of 
HAT-P-7b system. Using ellipsoidal variations we calculate the planetary mass from the photometric LC itself, without any RV measurments. 
Our technique will benefit from future space missions such as PLATO and the James Webb Space Telescope \citep{2009PASP..121..952D} with 
$\mathrm{RMS}~\sim~3\cdot 10^{-5}$ and $\lesssim 10^{-6}$, respectively.

\begin{acknowledgements}

The CoRoT space mission, launched on 2006 December 27, was developed and is operated by the CNES, with participation of the Science Programs of 
ESA, ESA's RSSD, Austria, Belgium, Brazil, Germany and Spain. This research has made use of NASA's Astrophysics Data System Bibliographic Services.

\end{acknowledgements}

\bibliographystyle{aa}
\bibliography{aa} 

\begin{thebibliography}{25}
\expandafter\ifx\csname natexlab\endcsname\relax\def\natexlab#1{#1}\fi

\bibitem[{{Alonso} {et~al.}(2008){Alonso}, {Barbieri}, {Rabus}, {Deeg},
  {Belmonte}, \& {Almenara}}]{2008A&A...487L...5A}
{Alonso}, R., {Barbieri}, M., {Rabus}, M., {et~al.} 2008, \aap, 487, L5

\bibitem[{{Borucki} {et~al.}(2011){Borucki}, {Koch}, {Basri}, {Batalha},
  {Brown}, {Bryson}, {Caldwell}, {Christensen-Dalsgaard}, {Cochran}, {DeVore},
  {Dunham}, {Gautier}, {Geary}, {Gilliland}, {Gould}, {Howell}, {Jenkins},
  {Latham}, {Lissauer}, {Marcy}, {Rowe}, {Sasselov}, {Boss}, {Charbonneau},
  {Ciardi}, {Doyle}, {Dupree}, {Ford}, {Fortney}, {Holman}, {Seager},
  {Steffen}, {Tarter}, {Welsh}, {Allen}, {Buchhave}, {Christiansen}, {Clarke},
  {Das}, {D{\'e}sert}, {Endl}, {Fabrycky}, {Fressin}, {Haas}, {Horch},
  {Howard}, {Isaacson}, {Kjeldsen}, {Kolodziejczak}, {Kulesa}, {Li}, {Lucas},
  {Machalek}, {McCarthy}, {MacQueen}, {Meibom}, {Miquel}, {Prsa}, {Quinn},
  {Quintana}, {Ragozzine}, {Sherry}, {Shporer}, {Tenenbaum}, {Torres},
  {Twicken}, {Van Cleve}, {Walkowicz}, {Witteborn}, \&
  {Still}}]{2011ApJ...736...19B}
{Borucki}, W.~J., {Koch}, D.~G., {Basri}, G., {et~al.} 2011, \apj, 736, 19

\bibitem[{{Borucki} {et~al.}(1997){Borucki}, {Koch}, {Dunham}, \&
  {Jenkins}}]{1997ASPC..119..153B}
{Borucki}, W.~J., {Koch}, D.~G., {Dunham}, E.~W., \& {Jenkins}, J.~M. 1997, in
  Astronomical Society of the Pacific Conference Series, Vol. 119, Planets
  Beyond the Solar System and the Next Generation of Space Missions, ed.
  {D.~Soderblom}, 153--+

\bibitem[{{Claret}(2004)}]{2004A&A...428.1001C}
{Claret}, A. 2004, A\&A, 428, 1001

\bibitem[{{Claudi}(2010)}]{2010Ap&SS.328..319C}
{Claudi}, R. 2010, \apss, 328, 319

\bibitem[{{Costes} {et~al.}(2004){Costes}, {Bodin}, {Levacher}, \&
  {Auvergne}}]{2004ESASP.554..281C}
{Costes}, V., {Bodin}, P., {Levacher}, P., \& {Auvergne}, M. 2004, in ESA
  Special Publication, Vol. 554, 5th International Conference on Space Optics,
  ed. {B.~Warmbein}, 281--284

\bibitem[{{Cowan} \& {Agol}(2010)}]{2010arXiv1011.0428C}
{Cowan}, N.~B. \& {Agol}, E. 2010, ArXiv e-prints

\bibitem[{{Deleuil} {et~al.}(1997){Deleuil}, {Barge}, {Leger}, \&
  {Schneider}}]{1997ASPC..119..259D}
{Deleuil}, M., {Barge}, P., {Leger}, A., \& {Schneider}, J. 1997, in
  Astronomical Society of the Pacific Conference Series, Vol. 119, Planets
  Beyond the Solar System and the Next Generation of Space Missions, ed.
  {D.~Soderblom}, 259--+

\bibitem[{{Deming} {et~al.}(2009){Deming}, {Seager}, {Winn}, {Miller-Ricci},
  {Clampin}, {Lindler}, {Greene}, {Charbonneau}, {Laughlin}, {Ricker},
  {Latham}, \& {Ennico}}]{2009PASP..121..952D}
{Deming}, D., {Seager}, S., {Winn}, J., {et~al.} 2009, \pasp, 121, 952

\bibitem[{{Faigler} \& {Mazeh}(2011)}]{2011MNRAS.415.3921F}
{Faigler}, S. \& {Mazeh}, T. 2011, \mnras, 415, 3921

\bibitem[{{Ford} {et~al.}(2008){Ford}, {Quinn}, \&
  {Veras}}]{2008ApJ...678.1407F}
{Ford}, E.~B., {Quinn}, S.~N., \& {Veras}, D. 2008, \apj, 678, 1407

\bibitem[{{Hansen}(2008)}]{2008ApJS..179..484H}
{Hansen}, B.~M.~S. 2008, \apjs, 179, 484

\bibitem[{{Kane} \& {Gelino}(2010)}]{2010ApJ...724..818K}
{Kane}, S.~R. \& {Gelino}, D.~M. 2010, \apj, 724, 818

\bibitem[{{Koch} {et~al.}(2010){Koch}, {Borucki}, {Basri}, {Batalha}, {Brown},
  {Caldwell}, {Christensen-Dalsgaard}, {Cochran}, {DeVore}, {Dunham},
  {Gautier}, {Geary}, {Gilliland}, {Gould}, {Jenkins}, {Kondo}, {Latham},
  {Lissauer}, {Marcy}, {Monet}, {Sasselov}, {Boss}, {Brownlee}, {Caldwell},
  {Dupree}, {Howell}, {Kjeldsen}, {Meibom}, {Morrison}, {Owen}, {Reitsema},
  {Tarter}, {Bryson}, {Dotson}, {Gazis}, {Haas}, {Kolodziejczak}, {Rowe}, {Van
  Cleve}, {Allen}, {Chandrasekaran}, {Clarke}, {Li}, {Quintana}, {Tenenbaum},
  {Twicken}, \& {Wu}}]{2010ApJ...713L..79K}
{Koch}, D.~G., {Borucki}, W.~J., {Basri}, G., {et~al.} 2010, \apjl, 713, L79

\bibitem[{{Loeb} \& {Gaudi}(2003)}]{2003ApJ...588L.117L}
{Loeb}, A. \& {Gaudi}, B.~S. 2003, \apjl, 588, L117

\bibitem[{{Mazeh} \& {Faigler}(2010)}]{2010A&A...521L..59M}
{Mazeh}, T. \& {Faigler}, S. 2010, \aap, 521, L59+

\bibitem[{{P{\'a}l} {et~al.}(2009{\natexlab{a}}){P{\'a}l}, {Bakos}, {Noyes}, \&
  {Torres}}]{2009IAUS..253..428P}
{P{\'a}l}, A., {Bakos}, G.~{\'A}., {Noyes}, R.~W., \& {Torres}, G.
  2009{\natexlab{a}}, in IAU Symposium, Vol. 253, IAU Symposium, 428--431

\bibitem[{{P{\'a}l} {et~al.}(2009{\natexlab{b}}){P{\'a}l}, {Bakos}, {Torres},
  {Noyes}, {Fischer}, {Johnson}, {Henry}, {Butler}, {Marcy}, {Howard},
  {Sip\H{o}cz}, {Latham}, \& {Esquerdo}}]{2009MNRAS.tmp.1781P}
{P{\'a}l}, A., {Bakos}, G.~{\'A}., {Torres}, G., {et~al.} 2009{\natexlab{b}},
  \mnras, 1781

\bibitem[{{Queloz} {et~al.}(2009){Queloz}, {Bouchy}, {Moutou}, {Hatzes},
  {H{\'e}brard}, {Alonso}, {Auvergne}, {Baglin}, {Barbieri}, {Barge}, {Benz},
  {Bord{\'e}}, {Deeg}, {Deleuil}, {Dvorak}, {Erikson}, {Ferraz Mello},
  {Fridlund}, {Gandolfi}, {Gillon}, {Guenther}, {Guillot}, {Jorda}, {Hartmann},
  {Lammer}, {L{\'e}ger}, {Llebaria}, {Lovis}, {Magain}, {Mayor}, {Mazeh},
  {Ollivier}, {P{\"a}tzold}, {Pepe}, {Rauer}, {Rouan}, {Schneider},
  {Segransan}, {Udry}, \& {Wuchterl}}]{2009A&A...506..303Q}
{Queloz}, D., {Bouchy}, F., {Moutou}, C., {et~al.} 2009, A\&A, 506, 303

\bibitem[{{Russell}(1916)}]{1916ApJ....43..173R}
{Russell}, H.~N. 1916, \apj, 43, 173

\bibitem[{{Seager} \& {Mall{\'e}n-Ornelas}(2003)}]{2003ApJ...585.1038S}
{Seager}, S. \& {Mall{\'e}n-Ornelas}, G. 2003, \apj, 585, 1038

\bibitem[{{Sozzetti} {et~al.}(2007){Sozzetti}, {Torres}, {Charbonneau},
  {Latham}, {Holman}, {Winn}, {Laird}, \& {O'Donovan}}]{2007ApJ...664.1190S}
{Sozzetti}, A., {Torres}, G., {Charbonneau}, D., {et~al.} 2007, \apj, 664, 1190

\bibitem[{{Sudarsky} {et~al.}(2000){Sudarsky}, {Burrows}, \&
  {Pinto}}]{2000ApJ...538..885S}
{Sudarsky}, D., {Burrows}, A., \& {Pinto}, P. 2000, \apj, 538, 885

\bibitem[{{von Zeipel}(1924)}]{1924MNRAS..84..665V}
{von Zeipel}, H. 1924, \mnras, 84, 665

\bibitem[{{Welsh} {et~al.}(2010){Welsh}, {Orosz}, {Seager}, {Fortney},
  {Jenkins}, {Rowe}, {Koch}, \& {Borucki}}]{2010ApJ...713L.145W}
{Welsh}, W.~F., {Orosz}, J.~A., {Seager}, S., {et~al.} 2010, \apjl, 713, L145

\end{thebibliography}

\end{document}